\documentclass[12pt,a4paper]{article}
\usepackage[latin1]{inputenc}
\usepackage{amsmath}
\usepackage{placeins}
\usepackage{makecell}
\usepackage{url}
\usepackage{natbib}
\usepackage{ulem}
\usepackage{color,xcolor,setspace,geometry}
\setcitestyle{authoryear,open={(},close={)}}
\usepackage{cite}
\usepackage{amsfonts}
\usepackage{tikz}
\usepackage{dsfont}
\usetikzlibrary{shapes}
\usetikzlibrary{plotmarks}
\usetikzlibrary{tikzmark,matrix,calc}
\usetikzlibrary{arrows.meta,calc,decorations.markings,math,arrows.meta}
\usepackage{amssymb}
\usepackage{multirow}
\usepackage{graphicx}
\begin{document}
\begin{spacing}{1.35}	

\title{A regularized hidden Markov model for analyzing the ``hot shoe'' in football}

\author{
Marius \"Otting\thanks{Department of Business Administration and Economics, Bielefeld University, Bielefeld, Germany},
Andreas Groll\thanks{Department of Statistics, TU Dortmund University, Dortmund, Germany} \\[1em] 
}

\date{}
\maketitle

\begin{abstract} 
Although academic research on the ``hot hand'' effect (in particular, in sports, especially in basketball) has been going on for more than 30 years, it still remains a central question in different areas of research whether such an effect exists.
In this contribution, we investigate the potential occurrence of a ``hot shoe'' effect for the performance of penalty takers in football based on data from the German Bundesliga. For this purpose, we consider hidden Markov models (HMMs) to model the (latent) forms of players. 
To further account for individual heterogeneity of the penalty taker as well as the opponent's goalkeeper, player-specific abilities are incorporated in the model formulation together with a LASSO penalty. 
Our results suggest states which can be tied to different forms of players, thus providing evidence for the hot shoe effect, and shed some light on exceptionally well-performing goalkeepers, which are of potential interest to managers and sports fans.
\end{abstract}

\section{Introduction}

In sports, the performance of players is frequently discussed by fans and journalists. An often discussed phenomenon in several sports is the ``hot hand'', meaning that players may enter a state where they experience extraordinary success.
For example, the former German football player Gerd M\"uller potentially was in a ``hot'' state when scoring 11 penalties in a row between 1975 and 1976. However, with 3 penalties missed in a row earlier in 1971, he potentially was in a ``cold'' state when taking these penalty kicks. 

Academic research on the hot hand started by \citet{gilovich1985hot}. 
In their seminal paper, they analyzed basketball free-throw data and provided 
no evidence for the hot hand, arguing that people tend to belief in the hot hand due to memory bias. 
In the past decade, however, some studies provided evidence for the hot hand 
while others failed to find such an effect (see \citealp{bar2006twenty}, for a review). 
Hence, the existence of a hot hand effect in different sports still remains an open question.

In our analysis, we investigate a potential ``hot shoe'' effect of penalty takers in the German Bundesliga. Our data set comprises all taken penalties in the Bundesliga from the first season (1963/64) until season 2016/17, totaling in $n = 3,482$ observations. Specifically, to explicitly account for the underlying (latent) form of a player, we consider hidden Markov models (HMMs) to investigate a potential hot shoe effect. Using HMMs to investigate the hot hand was first done by \citet{albert1993statistical} for an analysis in baseball, but also more recently by \citet{green2017hot} who also analyze data from baseball and by \citet{otting2018hot} who analyse data from darts. 

There are several potential confounding factors when analyzing the 
outcome of penalty kicks, such as the score of the match and the abilities 
of the two involved players, i.e.\ the penalty taker and the opposing team's 
goal keeper. Accounting for these factors leads to a large number of covariates, 
some of them also exhibiting a noteworthy amount of correlation/multicollinearity,
which makes model fitting and interpretation of parameters difficult. 
Hence, sparser models are desirable. To tackle these problems,  
variable selection is performed here by applying a LASSO penalization 
approach (see \citealp{Tibshirani:96}). 
Our results suggest clear evidence for two different states of penalty 
takers, which can be tied to a cold and a hot state. In addition, the results 
shed some light on exceptionally well-performing goalkeepers.

The remainder of the manuscript is structured as follows. The data on 
penalty kicks from the German Bundesliga is described in Section~\ref{chap:data}. 
In Section~\ref{chap:methods} the considered methodology is presented, namely a LASSO penalization technique for HMMs. 
The proposed approach is further investigated  in a short simulation study in 
Section~\ref{chap:simulation} and the results of our hot shoe analysis on German Bundesliga data
are presented in Section~\ref{chap:results}. Finally, we discuss the results and conclude in Section~\ref{chap:concl}.

\section{Data}\label{chap:data}

The considered data set comprises all taken penalty kicks in the German 
Bundesliga from its first season 1963/1964 until the end of the season 2016/2017. Parts of the data have already been used in \citet{bornkamp2009penalty}.
In the analysis, we include all players who took at least 5 penalty kicks 
during the time period considered, resulting in $n = 3,482$ penalty kicks taken by
310 different players. For these penalty kicks considered, 327 different 
goalkeepers were involved. The resulting variable of interest is a binary variable
indicating whether the player converted 
the penalty kick or not. Hence, we consider binary time series 
$\{ y_{p,t} \}_{t=1,\ldots,T_{p}}$, with $T_p$ denoting the total number of 
penalties taken by player $p$, 
indicating whether player $p$ scored the
penalty at attempt $t$, i.e.:

$$ 
y_{p,t} = 
\begin{cases}
1, & \text{if the $t$--th penalty kick is converted;} \\
0, & \text{otherwise.}
\end{cases}
$$
Since several other factors potentially affect the outcome of a penalty kick 
(such as the score of the match), we consider further covariates. For the 
choice of covariates, we follow \citet{dohmen2008professionals}, who analyzed 
the effect of pressure when taking penalty kicks and, hence, accounts for 
different potential confounders. These additional covariates include a 
dummy indicating whether the match was played at home, the matchday, 
the minute where the penalty was taken, the experience of both the penalty 
taker and the goalkeeper (quantified by the number of years the player played for a professional team) and the categorized score difference, with categories 
more than 2 goals behind, 2 goals behind, 1 goal behind, 1 goal ahead, 2 goals 
ahead or more than 2 goals ahead. Since the effect of the score might depend 
on the minute of the match, we further include interaction terms between the 
categories of the score difference and the minute. To consider rule changes for 
penalty kicks (see \citealp{dohmen2008professionals}, for more details), we 
include dummy variables for different time intervals (season 
1985/86 and before, between  season 1986/87 and  season 1995/96,  season 1996/1997, 
and from season 1997/1998 up to season 2016/2017).
Table \ref{tab:descriptives} summarizes descriptive statistics for all 
metric covariates considered as well as for $\{ y_{p,t} \}$.

\begin{table}[!htbp] \centering 
\caption{Descriptive statistics.} 
  \label{tab:descriptives} 
\begin{tabular}{@{\extracolsep{5pt}}lccccc} 
\\[-1.8ex]\hline 
\hline \\[-1.8ex] 
 & \multicolumn{1}{c}{Mean} & \multicolumn{1}{c}{St.\ Dev.} & \multicolumn{1}{c}{Min.} & \multicolumn{1}{c}{Max.} \\ 
\hline \\[-1.8ex] 
successful penalty &  0.780 & 0.414 & 0 & 1 \\ 
matchday &  -- & -- & 1 & 38 \\ 
home &  0.316 & 0.465 & 0 & 1 \\ 
experience (penalty taker) &  6.323 & 3.793 & 0 & 19 \\ 
experience (goalkeeper) &  5.343 & 4.187 & 0 & 19 \\ 
minute &  51.92 & 24.91 & 1 & 90 \\ 
\hline \\[-1.8ex] 
\end{tabular} 
\end{table} 
 
\begin{figure}[h ]
\centering
\includegraphics[scale=0.85]{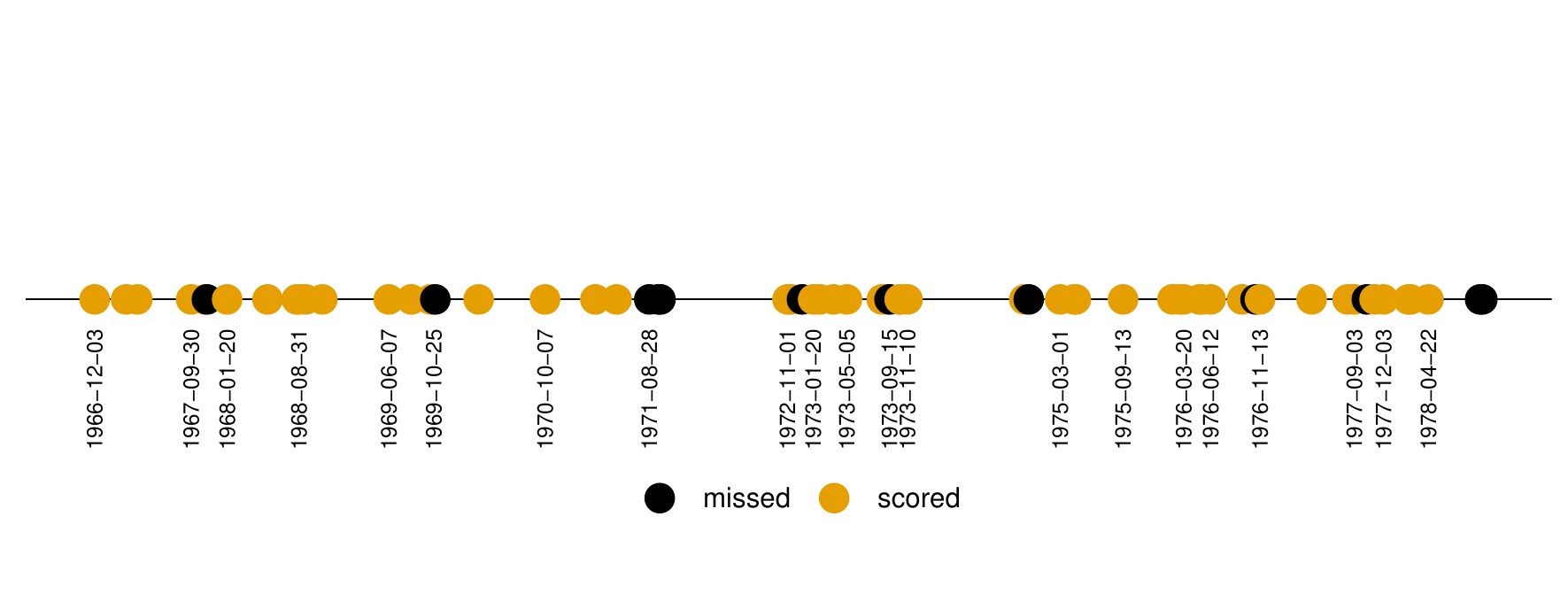}
\caption{Penalty history over time of the player Gerd M\"uller for the time period
from 1964 until 1979; a successful penalty is shown in yellow, a failure in black.} 
\label{fig:data}
\end{figure}

Finally, to explicitly account for player-specific characteristics, we include 
intercepts for all penalty takers as well as for all goalkeepers considered in our sample. 
These parameters can be interpreted as the players penalty abilities (i.e., the penalty shooting skill
for the penalty taker and the (negative) penalty saving skill for the goalkeeper).
To illustrate the typical structure of our data, an example time series from our sample of the famous German attacker Gerd M\"uller, who played in the Bundesliga for Bayern 
Munich from 1964 until 1979, is shown in Figure \ref{fig:data}. The corresponding part in the data set is shown in Tables \ref{tab:design1} and \ref{tab:design2}.

\begin{table}[ht]
\centering
\caption{Part of the data set corresponding to the metric covariates.}
\label{tab:design1}
\scalebox{0.9}{
\begin{tabular}{cccccccc}
  \hline
\thead{\textbf{player}} & \thead{\textbf{successful} \\ \textbf{penalty}} & \thead{\textbf{matchday}} & \thead{\textbf{home}} & \thead{\textbf{experience} \\ \textbf{(penalty taker)}} & \thead{\textbf{experience} \\ \textbf{(goalkeeper)}} & \thead{\textbf{minute}} & \thead{$\pmb{\cdots}$} \\ 
  \hline
  $\vdots$ & $\vdots$ & $\vdots$ & $\vdots$ & $\vdots$ & $\vdots$ & $\vdots$ & $\cdots$ \\
Gerd M\"uller & 1 & 15 & 0 & 1 & 3 & 90 & $\cdots$ \\ 
  Gerd M\"uller & 1 & 25 & 0 & 1 & 3 & 81 & $\cdots$ \\ 
    $\vdots$ & $\vdots$ & $\vdots$ & $\vdots$ & $\vdots$ & $\vdots$ & $\vdots$ & $\cdots$ \\
  Gerd M\"uller & 0 & 7 & 0 & 13 & 2 & 37 & $\cdots$ \\ 
  Gerd M\"uller & 0 & 8 & 1 & 13 & 8 & 68 & $\cdots$ \\ 
    $\vdots$ & $\vdots$ & $\vdots$ & $\vdots$ & $\vdots$ & $\vdots$ & $\vdots$ & $\ddots$ \\
   \hline
\end{tabular}}
\end{table}

\begin{table}[ht]
\centering
\caption{Part of the data set corresponding to the player- and goalkeeper specific effects.}
\label{tab:design2}
\scalebox{0.85}{
\begin{tabular}{ccccccccc}
  \hline
\thead{\textbf{Hans} \\ \textbf{M\"uller} \\  \textbf{(player)}} & \thead{\textbf{Gerd} \\ \textbf{M\"uller} \\ \textbf{(player)}} & \thead{\textbf{Ludwig} \\ \textbf{M\"uller} \\ \textbf{(player)}} &  \thead{$\pmb{\cdots}$} & 
\thead{\textbf{G\"unter} \\ \textbf{Bernard} \\ \textbf{(goalkeeper)}} &
\thead{\textbf{Wolfgang} \\ \textbf{Schnarr} \\ \textbf{(goalkeeper)}} &
\thead{$\pmb{\cdots}$} &
\thead{\textbf{Dieter} \\ \textbf{Burdenski} \\ \textbf{(goalkeeper)}} &   
\thead{\textbf{Wolfgang} \\ \textbf{Kneib} \\ \textbf{(goalkeeper)}} \\ 
  \hline
  $\vdots$ & $\vdots$ & $\vdots$ & $\cdots$ & $\vdots$ & $\vdots$ & $\cdots$ & $\vdots$ & $\vdots$ \\
0 & 1 & 0 & $\cdots$ & 0 & 1 & $\cdots$ & 0 & 0 \\ 
  0 & 1 & 0 & $\cdots$ & 1  & 0 & $\cdots$ & 0 & 0 \\ 
    $\vdots$ & $\vdots$ & $\vdots$ & $\cdots$ & $\vdots$ & $\vdots$ & $\cdots$ & $\vdots$ & $\vdots$ \\
  0 & 1 & 0 & $\cdots$ & 0 &  0 & $\cdots$ & 0 & 1 \\ 
  0 & 1 & 0 & $\cdots$ & 0 & 0 & $\cdots$ & 1 & 0 \\ 
    $\vdots$ & $\vdots$ & $\vdots$ & $\cdots$ & $\vdots$ & $\vdots$ & $\cdots$ & $\vdots$ & $\ddots$ \\
   \hline
\end{tabular}}
\end{table}

\section{Methods}\label{chap:methods}
Figure \ref{fig:data} indicates that there are phases in the career of Gerd M\"uller 
where he scored several penalty kicks in a row, e.g.\ between 1975 and 1976 as already
mentioned in the introduction. At some parts of his career, however, successful penalty kicks 
were occasionally followed by one or more missed penalty kicks. To
explicitly account for such phases we consider HMMs, where the latent state 
process can be interpreted as the underlying varying form of a player. 
Moreover, \citet{stone2012measurement} argues that HMMs are more suitable 
for analyzing the hot hand than analyzing serial correlation of outcomes, since the latter 
mentioned outcomes are only noisy measures of the underlying (latent) form of a player.

\subsection{Hidden Markov models}\label{sec:hmm}

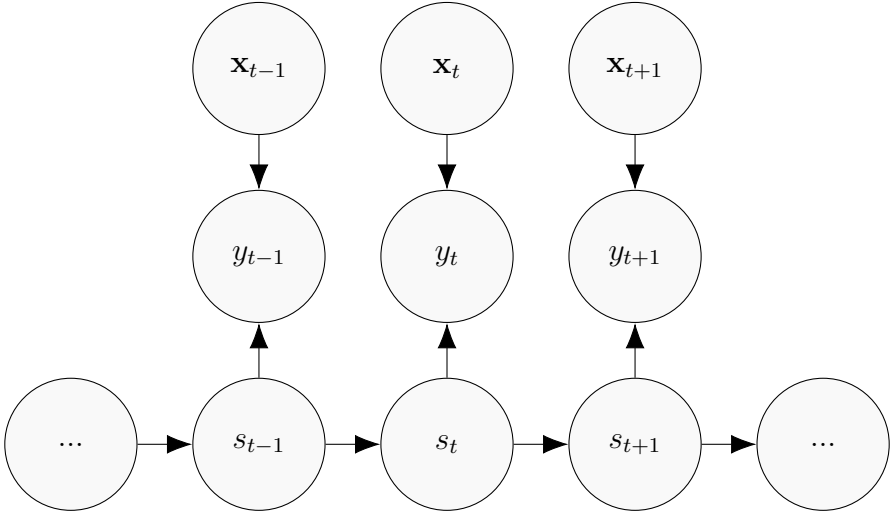
\begin{figure}[h!]
    \centering
	\begin{tikzpicture}
	\node[circle,draw=black, fill=gray!5, inner sep=0pt, minimum size=50pt] (A) at (2, -5) {$s_{t-1}$};
	\node[circle,draw=black, fill=gray!5, inner sep=0pt, minimum size=50pt] (A1) at (-0.5, -5) {...};
	\node[circle,draw=black, fill=gray!5, inner sep=0pt, minimum size=50pt] (B) at (4.5, -5) {$s_{t}$};
	\node[circle,draw=black, fill=gray!5, inner sep=0pt, minimum size=50pt] (C) at (7, -5) {$s_{t+1}$};
	\node[circle,draw=black, fill=gray!5, inner sep=0pt, minimum size=50pt] (C1) at (9.5, -5) {...};
	\node[circle,draw=black, fill=gray!5, inner sep=0pt, minimum size=50pt] (Y1) at (2, -2.5) {$y_{t-1}$};
	\node[circle,draw=black, fill=gray!5, inner sep=0pt, minimum size=50pt] (Y2) at (4.5, -2.5) {$y_{t}$};
	\node[circle,draw=black, fill=gray!5, inner sep=0pt, minimum size=50pt] (Y3) at (7, -2.5) {$y_{t+1}$};
	\node[circle,draw=black, fill=gray!5, inner sep=0pt, minimum size=50pt] (X3) at (7, 0) {$\mathbf{x}_{t+1}$};
	\node[circle,draw=black, fill=gray!5, inner sep=0pt, minimum size=50pt] (X2) at (4.5, 0) {$\mathbf{x}_{t}$};
	\node[circle,draw=black, fill=gray!5, inner sep=0pt, minimum size=50pt] (X1) at (2, 0) {$\mathbf{x}_{t-1}$};
	\draw[-{Latex[scale=2]}] (A)--(B);
	\draw[-{Latex[scale=2]}] (B)--(C);
	\draw[-{Latex[scale=2]}] (A1)--(A);
	\draw[-{Latex[scale=2]}] (C)--(C1);
	\draw[-{Latex[scale=2]}] (A)--(Y1);
	\draw[-{Latex[scale=2]}] (B)--(Y2);
	\draw[-{Latex[scale=2]}] (C)--(Y3);,
	\draw[-{Latex[scale=2]}] (X1)--(Y1);
	\draw[-{Latex[scale=2]}] (X2)--(Y2);
	\draw[-{Latex[scale=2]}] (X3)--(Y3);
	\end{tikzpicture}
\caption{Graphical structure of the HMM considered. Each observation $y_{t}$ is 
assumed to be generated by one of $N$ distributions according to the state process 
$s_{t}$, which serves for the underlying form of a player. 
In addition, covariates $\mathbf{x}_{t}$ are assumed to affect $y_t$.}
\label{fig:HMM}
\end{figure}

In HMMs, the observations $y_{p,t}$ are assumed to be driven by an underlying state process 
$s_{p,t}$, in a sense that the $y_{p,t}$ are generated by one of $N$ distributions
according to the Markov chain. In our application, the state process $s_{p,t}$ 
serves for the underlying varying form of a player. For notational simplicity, 
we drop the player-specific subscript $p$ in the following. Switching between 
the states is taken into account by the transition probability matrix (t.p.m.) 
$\boldsymbol{\Gamma} = (\gamma_{ij})$, with $\gamma_{ij} =  \Pr(s_{t} = j | s_{t-1} = i),\, i,j = 1,\ldots,N$. 
We further allow for additional covariates at time $t$, $\boldsymbol{x}_t = (x_{1t}, \ldots, x_{Kt})'$, 
each of which assumed to have the same effect in each state, 
whereas the intercept is assumed to vary across the states, 
leading to the following linear state-dependent predictor:

$$
\eta^{(s_{t})} = \beta_0^{(s_t)} + \beta_1 x_{1t} + \ldots + \beta_k x_{Kt}.
$$ 
In fact, this is a simple Markov-switching regression model, where only the intercept varies across the states (see, e.g., \citealp{goldfeld1973markov}). The dependence structure of the HMM considered
is shown in Figure \ref{fig:HMM}.

For our response variable $y_t$, indicating whether the penalty attempt $t$ 
was successful or not, we assume $y_{t} \sim \text{Bern}(\pi_{t})$ and 
link $\pi_t$ to our state-dependent linear predictor $\eta_t$ using the logit link function, i.e.\ 
$\text{logit}(\pi_t) = \eta^{(s_{t})}$. Defining an $N \times N$ diagonal matrix 
$\boldsymbol{P}(y_t)$ with $i$--th diagonal element equal to
$\Pr(y_t|s_t = i)$ and assuming that the initial distribution $\boldsymbol{\delta}$ 
of a player is equal to the stationary distribution, i.e.\ the solution to 
$\boldsymbol{\Gamma} \boldsymbol{\delta} = \boldsymbol{\delta}$, the likelihood for a single player $p$ is given by

\begin{equation*}
L_p(\pmb{\alpha}) = \boldsymbol{\delta} \mathbf{P}(y_{p1}) \mathbf{\Gamma}\mathbf{P}(y_{p2}) \dots \mathbf{\Gamma}\mathbf{P}(y_{pT_p}) \mathbf{1}\,,
\end{equation*}
with column vector $\mathbf{1}=(1,\ldots,1)' \in \mathbb{R}^N$ 
(see \citealp{zucchini2016hidden})
and parameter vector $\pmb{\alpha}=(\gamma_{11},\gamma_{12},\ldots,\gamma_{1N},\ldots,\gamma_{NN},\beta_0^{(1)},\ldots,\beta_0^{(N)},
\beta_1,\ldots,\beta_k)'$ collecting all unknown parameters. Specifically, formulating the likelihood as above amounts to running the forward algorithm, which allows to calculate the 
likelihood recursively at computational cost $\mathcal{O}(TN^2)$ only, thus rendering
numerical maximization of the likelihood feasible \citep{zucchini2016hidden}. To obtain the full 
likelihood for all 310 players considered in the sample, we assume independence between
the individual players such that the likelihood is calculated by the 
product of the individual likelihoods of the players:

\begin{equation*}
L(\pmb{\alpha}) = \prod_{p=1}^{310} L_p(\pmb{\alpha}) = \prod_{p=1}^{310} \boldsymbol{\delta} \mathbf{P}(y_{p1}) \mathbf{\Gamma}\mathbf{P}(y_{p2}) \dots \mathbf{\Gamma}\mathbf{P}(y_{pT_p}) \mathbf{1}.
\end{equation*}
For our analysis of a potential hot shoe effect, we initially select $N=2$ states, which potentially are aligned to a ``hot'' and a ``cold'' state, i.e.\ states with superior and poor performance, respectively. The parameter vector $\pmb{\alpha}$, hence, reduces to $\pmb{\alpha}=(\gamma_{11}, \gamma_{12}, \gamma_{21}, \gamma_{22},\beta_0^{(1)}, \beta_0^{(2)}, \beta_1, \ldots, \beta_k)'$. The choice of $N$ will be further discussed in Chapter \ref{chap:results}.

Parameter estimation is done by maximizing the likelihood numerically using \texttt{nlm()} in R \citep{rcoreteam}. However, if we consider all covariates introduced in our model formulation from Section~\ref{chap:data}, the model gets rather complex, is hard to interpret and multicollinearity issues might occur. Hence, we propose to employ a penalized likelihood approach based on a LASSO penalty, 
which is described in the next section.

\subsection{Variable selection by the LASSO}\label{sec:lasso}

In order to obtain a sparse and interpretable model, 
the estimation of the covariate effects will be performed by a
regularized estimation approach. The idea is to first set up a model with a rather large number
of possibly influential variables (in particular, with regard to the player-specific ability parameters)
and then to regularize the effect of the single covariates. This way, the variance of the parameter 
estimates is diminished and, hence, usually lower prediction 
error is achieved than with the unregularized maximum likelihood (ML) estimator. 
The basic concept of regularization is to maximize a penalized version of the likelihood 
$\ell(\boldsymbol{\alpha}) = \log\left(L(\boldsymbol{\alpha})\right)$. 
More precisely, one maximizes the penalized log-likelihood
\begin{equation}
\ell_{\text{pen}}(\boldsymbol{\alpha}) = \log\left(L(\boldsymbol{\alpha})\right) - \lambda \sum_{k=1}^{K} |\beta_k|\,,
\label{eq:pen_likelihood}
\end{equation}
where $\lambda$ represents a tuning parameter, which controls the 
strength of the penalization. The optimal value for this tuning parameter 
has to be chosen either by cross-validation or suitable model selection criteria. The latter
usually constitute a compromise between the model fit (e.g., in terms of the likelihood) 
and the complexity of the model. Frequently used are the Akaike information criterion (AIC; \citealp{Akaike:73}) or Bayesian information criterion (BIC; \citealp{Schwarz:78}). In the context of LASSO, the effective degrees of freedom for the AIC and BIC are estimated as the number of non-zero coefficients (see \citealp{zou2007degrees}). Since our longitudinal data structure with multiple short time series from 310 individuals renders cross-validation rather difficult, we select the tuning parameter $\lambda$ in the following by information criteria. 

Note that in contrast to the ridge penalty, which penalizes 
the squared coefficients and shrinks them towards zero  (see \citealp{HoeKen:70}), 
the LASSO penalty on the absolute values of the coefficients, first proposed by \citet{Tibshirani:96},
can set coefficients to exactly zero and, hence, enforces variable selection.

Another advantage of the employed penalization is the way correlated predictors are treated.
For example, if two (or more) predictors are highly correlated, parameter estimates are stabilized by the penalization.
In such scenarios, the LASSO penalty tends to include only one of the predictors and 
only includes a second predictor if it entails additional information for the response variable. 
Therefore, if several variables possibly contain information on the outcome of the penalty,
they can be used simultaneously.

To fully incorporate the LASSO penalty in our setting, the non-differentiable $L_1$ norm $|\beta_k|$ in Eq.\ (\ref{eq:pen_likelihood}) is approximated as suggested by \citet{oelker2017uniform}. Specifically, the $L_1$ norm is approximated by $\sqrt{(\beta_k + c)^2}$, where $c$ is a small positive number (say c = $10^{-5}$). With the approximation of the penalty, the corresponding likelihood is still maximized numerically using  \texttt{nlm()} in R as denoted above.

In the simulation study from the subsequent section, we also investigate a relaxed LASSO-type 
version of our fitting scheme. The relaxed LASSO \citep{Mein:2007}
is known to often produce sparser models with equal or lower prediction loss than the regular LASSO.
To be more precise, for each value of the tuning parameter $\lambda$, in a final 
step we fit an (unregularized) model that includes only the variables
corresponding to the non-zero parameters of the preceding LASSO estimates.

\section{A short simulation study}\label{chap:simulation}

We consider a simulation scenario similar to our real-data application, with a Bernoulli-distributed response variable, an underlying 2-state Markov chain and 50
covariates, 47 of which being noise covariates:
$$
y_t \sim \text{Bern}(\pi^{(s_{t})}),
$$ 
with
$$
\text{logit}(\pi^{(s_{t})}) = \eta^{(s_{t})} = \beta_0^{(s_t)} + 0.5 \cdot x_{1t} + 0.7 \cdot x_{2t} -0.8 \cdot x_{3t} + \sum_{j=4}^{47} 0\cdot x_{jt}\,.
$$
We further set $\beta_0^{(1)} = \text{logit}^{-1}(0.75), \beta_0^{(2)} = \text{logit}^{-1}(0.35)$ and

\begin{align*} 
\pmb{\Gamma} = 
\begin{pmatrix}
0.9 & 0.1 \\
0.1 & 0.9  \\
\end{pmatrix}.
\end{align*} 
The covariate values were drawn independently from a uniform distribution within the interval $[-2, 2]$. The interval boundaries as well as the
corresponding effects $\beta_1, \beta_2$ and $\beta_3$ were chosen such that reasonable values for the response are obtained (i.e., moderate proportions of ones and zeros). We conduct 100 simulation runs,
in each run generating $T = 5100$ observations $y_t, t = 1, \ldots, 5100,$ from the model specified above, with the sample size being about the same size as for the real data application. Out of these 5100 simulated observations, the first 5000 are used for model fitting, whereas for the last 100 observations (which are denoted by $y_t^{\text{test}}$), the predictive performance of several different models is compared (see below).

For the choice of the tuning parameter $\lambda$, we consider a (logarithmic) grid of length 50, $\Lambda= \{5000, \ldots, 0.0001\}$. To compare the performance of the above 
described LASSO-type estimation, we consider the following five fitting schemes:
\begin{itemize}
    \item HMM without penalization (i.e., with $\lambda = 0$) 
    \item LASSO-HMM with $\lambda$ selected by AIC 
    \item LASSO-HMM with $\lambda$ selected by BIC
    \item relaxed-LASSO-HMM with $\lambda$ selected by AIC 
    \item relaxed-LASSO-HMM with $\lambda$ selected by BIC
\end{itemize}
For all five methods considered, we calculate the mean squared error (MSE) of the 
coefficients $\beta_1, \ldots, \beta_{50}$:

$$
\text{MSE}_{\boldsymbol{\beta}} = \dfrac{1}{50} \sum_{k=1}^{50} (\hat{\beta}_k -\beta_k)^2.
$$
We also calculate the MSE for the state-dependent intercepts $\beta_0^{(1)}$ and $\beta_0^{(2)}$, and for the entries of the t.p.m., $\gamma_{11}$ and $\gamma_{22}$, which is done analogously to the MSE for $\beta_1, \ldots, \beta_{50}$ shown above.
To further compare the predictive performance of the fitting schemes considered, we predict the distribution for each of the 100 out-of-sample observations, i.e.\ the success probabilities $\hat{\pi}_{t}^{\text{pred}}$, and compare these to the 100 remaining simulated observations $y_t^{\text{test}}$. For that purpose, we calculate the Brier score and the average predicted probability, which are given as follows:

\begin{align*} 
\begin{split}
B & = \dfrac{1}{100} \sum_{t=1}^{100} (\hat{\pi}_{t}^{\text{pred}} - y_t^{\text{test}})^2 \\
A & = \dfrac{1}{100} \sum_{t=1}^{100} \left(\hat{\pi}_t^{pred} \mathds{1}_{\{ y_t^{test} = 1 \}} + (1 - \hat{\pi}_t^{pred}) \mathds{1}_{\{ y_t^{test} = 0 \}} \right),
\end{split}
\end{align*} 
with $\mathds{1}_{\{.\}}$ denoting the indicator function. For the Brier score $B$, more accurate predictions correspond to lower values, with the lowest possible value being 0. For the average predicted probability $A$, higher values correspond to more precise predictions. In addition, the average predicted probability can be directly interpreted as the probability for a correct prediction.

\begin{figure}[!htb]
\centering
\includegraphics[scale=0.69]{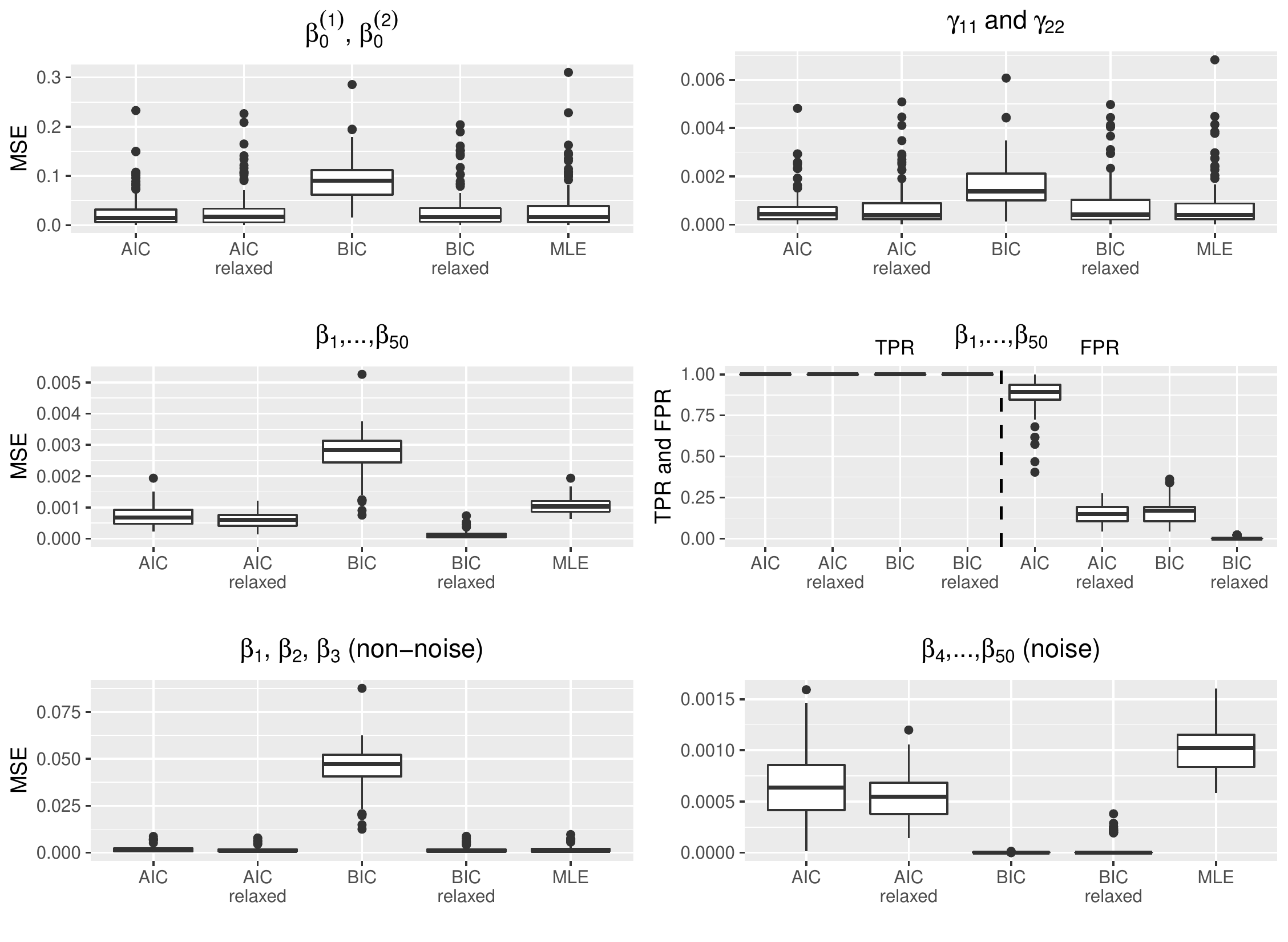}
\caption{Boxplots of the MSE, TPR and FPR obtained in 100 simulation runs. ``AIC'' and ``BIC'' denote the LASSO-HMM fitting scheme with $\lambda$ chosen by AIC and BIC, respectively. ``AIC relaxed'' and ``BIC relaxed" denote the relaxed-LASSO-HMM fitting scheme with $\lambda$ chosen by AIC and BIC, respectively. ``MLE'' denots the HMM without penalization.} 
\label{fig:simulation1}
\end{figure}

\begin{figure}[!htb]
\centering
\includegraphics[scale=0.775]{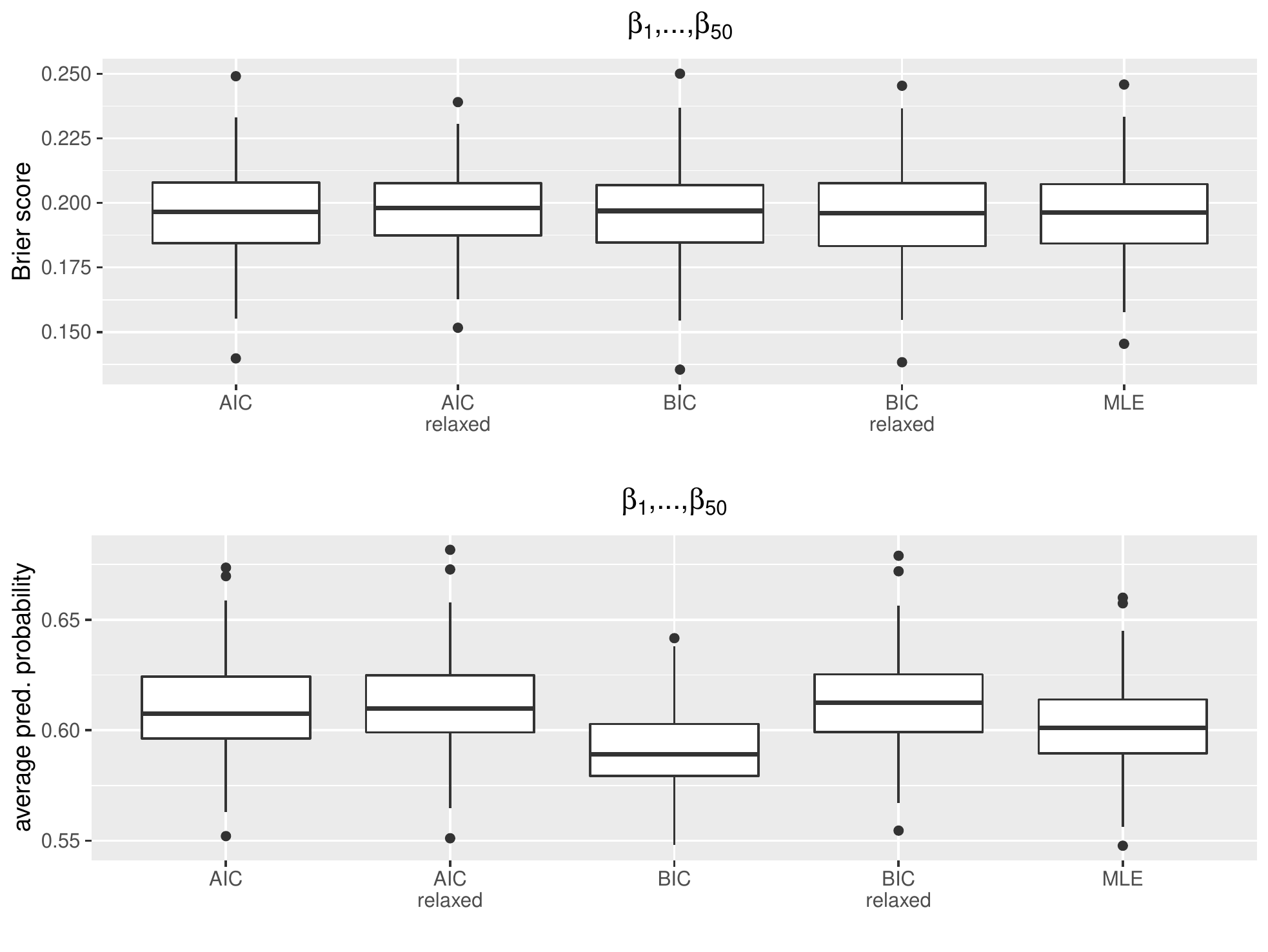}
\caption{Boxplots of the Brier score (top panel) and the average predicted probability (bottom panel)  obtained in 100 simulation runs.``AIC" and ``BIC" denote the LASSO-HMM fitting scheme with $\lambda$ chosen by AIC and BIC, respectively. ``AIC relaxed" and ``BIC relaxed" denote the relaxed-LASSO-HMM fitting scheme with $\lambda$ chosen by AIC and BIC, respectively. ``MLE" denots the HMM without penalization.} 
\label{fig:simulation2}
\end{figure}

The boxplots showing the MSEs over the 100 simulation runs, and the boxplots showing the Brier score and the average predicted probability for the predictive performance, are shown in Figures \ref{fig:simulation1} and \ref{fig:simulation2}, respectively. In both figures, the HMM without penalization is denoted by ``MLE", the LASSO-HMM with $\lambda$ selected by AIC and BIC are denoted by ``AIC" and ``BIC", respectively, and the relaxed-LASSO-HMM with $\lambda$ selected by AIC and BIC are denoted by ``AIC relaxed" and ``BIC relaxed", respectively. For the state-dependent intercepts $\beta_0^{(1)}$ and $\beta_0^{(2)}$, we see that the median MSE for the models with $\lambda$ chosen by the BIC is fairly high compared to all other models considered. A similar behaviour is observed for the entries of the t.p.m., $\gamma_{11}$ and $\gamma_{22}$. 

The middle row in Figure \ref{fig:simulation1} shows the MSE for $\beta_1, \ldots \beta_{50}$ as well as the corresponding true and false positive rates (denoted by TPR and FPR, respectively). The simulation results indicate that the non-noise covariates are detected by all models, whereas especially the LASSO-HMM with $\lambda$ selected by the AIC detects several noise covariates. A fairly low number of noise coefficients is selected by the relaxed-LASSO-HMM fitting scheme with $\lambda$ selected by the AIC and by the LASSO-HMM with $\lambda$ selected by the the BIC. The corresponding medians for the FPR are 0.149 and 0.170, respectively. The most promising results are given by the relaxed-LASSO-HMM with $\lambda$ chosen by the BIC. In 84 out of 100 simulations, no noise covariates were selected by this model.

The left plot in the middle row of Figure \ref{fig:simulation1} shows that the median MSE of the coefficients $\beta_j$ for the LASSO-HMM with $\lambda$ selected by the BIC is higher than the median MSE of all other models considered. This arises since the BIC tends to select a rather high $\lambda$, which can be seen from the median MSE separated for the noise and non-noise coefficients (last row of Figure \ref{fig:simulation1}). With the fairly high $\lambda$ chosen by the BIC, i.e.\ with more shrinkage involved, the MSE for the non-noise coefficients is rather large. At the same time, since only a few covariates are selected with a rather high $\lambda$, the MSE for the noise coefficients is very low.

For the predictive performance of the methods considered, we see that visually there is no clear difference in the Brier score between the models, which is shown in the top panel of Figure \ref{fig:simulation2}. The result for the average predicted probability -- shown in the bottom panel of Figure \ref{fig:simulation2} -- confirm that the LASSO-HMM with $\lambda$ chosen by the BIC and the HMM without penalization perform worse than the other models considered.

The results of the simulation study are very encouraging, with the LASSO penalty allowing for variable selection. The performance in terms of MSE, TPR and FPR suggest that the LASSO-HMM with $\lambda$ selected by the BIC performs worst, with the MSE being higher than for the HMM without penalization. However, the LASSO-HMM with $\lambda$ selected by the AIC as well as the relaxed-LASSO-HMM with $\lambda$ selected by AIC and BIC, respectively, (partly substantially) outperform the HMM without penalization in terms of MSE. The relaxed-LASSO-HMM with $\lambda$ selected by the BIC performs best in terms of MSE, TPR, FPR and the predictive performance. 
Finally, in all simulation runs the overall pattern was captured with regard to the true underlying state-dependent intercepts $\beta_0^{(1)}, \beta_0^{(2)}$ and diagonal entries of the t.p.m., i.e.\ $\gamma_{11}$ and $\gamma_{22}$. Fitting the LASSO-HMM and the relaxed-LASSO-HMM on the grid containing 50 different tuning parameters $\lambda$ took on average 47 minutes using a 3.4 GHz Intel\textsuperscript{\textcopyright} Core$^{\text{TM}}$ i7 CPU. This is remarkably fast, considering that both the LASSO-HMM and the relaxed-LASSO-HMM are fitted to the data for each value of $\lambda$, leading to 100 fitted models in total.

\FloatBarrier
\section{Results}\label{chap:results}
We now apply our LASSO-HMM approach to the German Bundesliga penalty data.
For the analysis of a potential hot shoe effect, we include all covariates from Section~\ref{chap:data} into the predictor and chose $N=2$ for the number of states, which can be interpreted as hot and cold states, respectively.\footnote{A psychological reason for being hot or cold may be a higher/lower level of self confidence.} This yields the following linear state-dependent predictor\footnote{Throughout this paper, all metric covariates are considered as linear. Since the main interest of this contribution is to investigate the LASSO penalty in HMMs, future research on the hot shoe could focus also on non-linear effects, for example for the matchday and the minute.}:
$$
\text{logit}(\pi^{(s_t)}) = \beta_0^{(s_t)}\, +\, \beta_1 \text{home}_{t}\, + \,\beta_2 \text{minute}_{t} \,+ \ldots + \,\beta_{100} \text{GerdMueller}_{t}\, + \ldots +\, \beta_{656}\text{WolfgangKneib}_{t}\,.
$$ 
Since the simulation study above indicates that the unpenalized maximum likelihood estimator is not appropriate for such a large number of covariates, we only consider the LASSO-type fitting schemes for the real data application. Specifically, since the relaxed-LASSO-HMM with $\lambda$ selected by the BIC showed the most promising results in the simulation, we focus on the results obtained by this fitting scheme, but we also present the results obtained by the two LASSO-HMM fitting schemes\footnote{The relaxed-LASSO-HMM with optimal $\lambda$ selected by the AIC yielded a rather unreasonable model, where almost all of the more than 600 covariates were selected and with partly unrealistically large corresponding estimated covariate effects, indicating some tendency of overfitting. For this reason, we excluded this model from the analysis.}.
The parameter estimates obtained (on the logit scale) indicate that the baseline level for scoring a penalty is higher in the model's state 1 than in state 2 ($\hat{\beta}_0^{(1)} = 1.422 > \hat{\beta}_0^{(2)} = -14.50$), thus indicating evidence for a hot shoe effect. State 1, hence, can be interpreted as a hot state, whereas state 2 refers to a cold state. In addition, with the t.p.m.\ estimated as

\begin{align*} 
\hat{\mathbf{\Gamma}} = 
\begin{pmatrix}
0.978 & 0.022 \\
0.680 & 0.320 \\
\end{pmatrix},
\end{align*} 
there is high persistence in state 1, i.e.\ in the hot state, whereas the cold state 2 is a transient state, where switching to state 1 is most likely. The stationary distribution as implied by the estimated t.p.m.\ is $\hat{\boldsymbol{\delta}} = (0.969, 0.031)$, i.e., according to the fitted model, players are in about 96.9\% of the time in the hot state and in about 3.1\% in the cold state. The diagonal elements of the t.p.m.\ for the other fitting schemes are obtained as $\hat{\gamma}_{11, \text{AIC}} = 0.989, \, \hat{\gamma}_{22, \text{AIC}} =  0.386$ and $\hat{\gamma}_{11, \text{BIC}} =  0.987, \, \hat{\gamma}_{22, \text{BIC}} = 0.368$, respectively. 
The corresponding state-dependent intercepts are obtained as $\hat{\beta}_{0, \text{AIC}}^{(1)} = -14.71, \, \hat{\beta}_{0, \text{AIC}}^{(2)} = 1.347$ and $\hat{\beta}_{0, \text{BIC}}^{(1)} = -18.83, \, \hat{\beta}_{0, \text{BIC}}^{(2)} = 1.360$, respectively. These results further confirm evidence for a hot shoe effect.

\begin{figure}[!htb]
\centering
\includegraphics[scale=0.75]{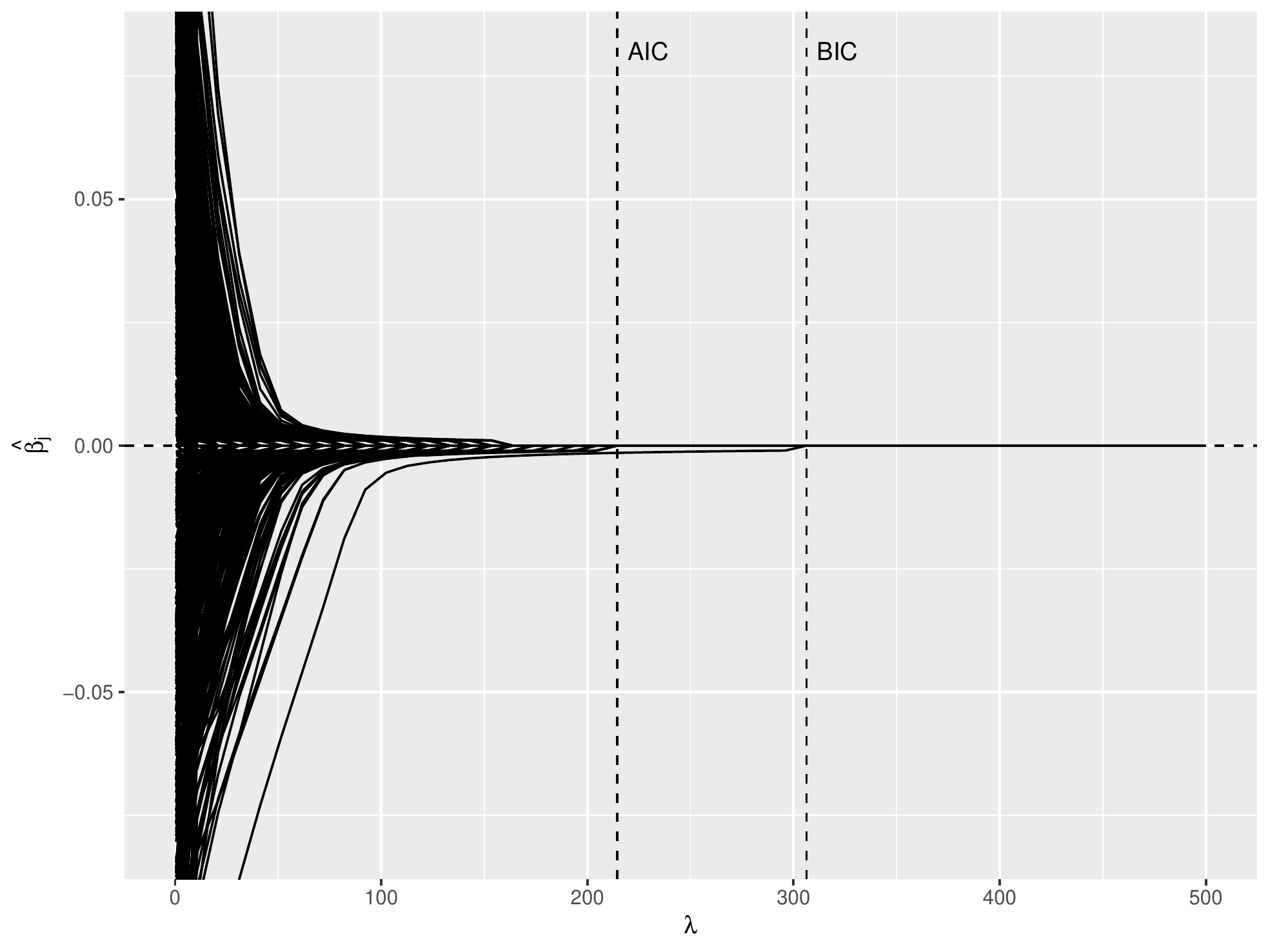}
\caption{Coefficient paths of all covariates considered in the LASSO-HMM models. The dashed lines indicate the optimal penalty parameters $\lambda$ selected by AIC and BIC, respectively. For the $\lambda$ selected by the BIC, no covariates are selected, whereas for the AIC one covariate is selected (see also Table \ref{tab:overview_effects}).
}
\label{fig:pathslasso}
\end{figure}

\begin{figure}[!htb]
\centering
\includegraphics[scale=0.75]{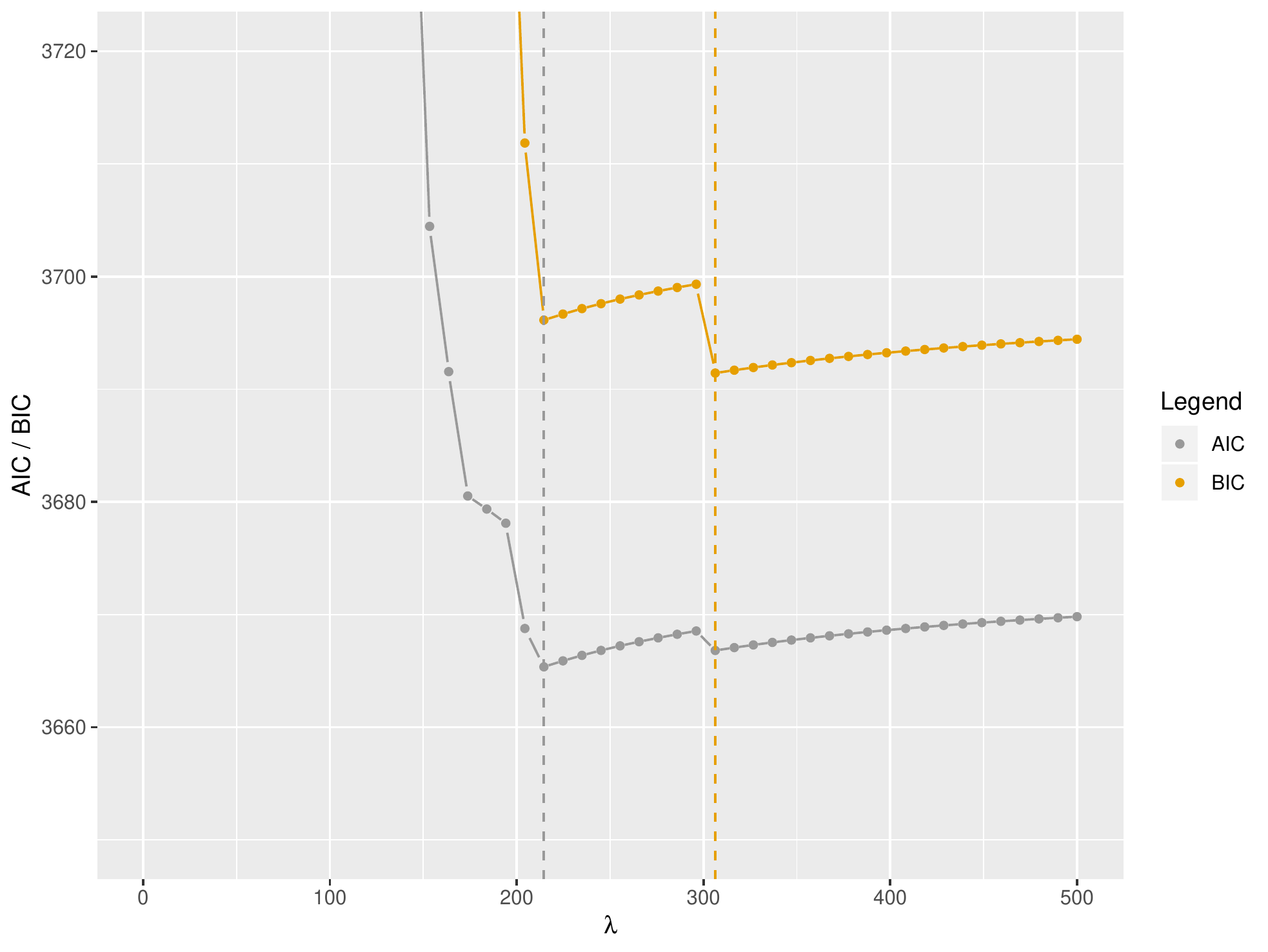}
\caption{Paths of AIC and BIC in the LASSO-HMM models. The vertical lines indicate the optimal penalty parameters $\lambda$ selected by AIC and BIC, respectively.
}
\label{fig:aicbic_fin}
\end{figure}

For the grid of potential tuning parameters $\lambda$, Figure \ref{fig:aicbic_fin} shows the progress of the AIC and BIC for the LASSO-HMM, indicating that the AIC selects a lower tuning parameter than the BIC. The corresponding coefficient paths of the LASSO-HMM approach together with the associated optimal tuning parameters selected by the AIC and BIC, respectively, are shown in Figure \ref{fig:pathslasso}.\footnote{We abstain from showing the coefficient paths plot for the relaxed LASSO-HMM model, because due to the unpenalized re-fit the paths look rather irregular.}  No covariates are selected by the LASSO-HMM with $\lambda$ selected by the BIC, whereas Jean-Marie Pfaff, a former goalkeeper of Bayern Munich, is selected by the LASSO-HMM with $\lambda$ selected by the AIC and by the relaxed-LASSO-HMM model based on BIC. The negative coefficient implies that the odds for scoring a penalty decrease if Jean-Marie Pfaff is the goalkeeper of the opponent's team. 
The coefficient is substantially larger in magnitude for the relaxed-LASSO model, since for this fitting schemes the model is re-fitted with $\lambda = 0$ on the set of selected coefficients from the first model fit (see above). An overview of the selected effects is given in Table \ref{tab:overview_effects}.

\begin{table}[ht]
\centering
\caption{Overview of selected players and goalkeepers by all models considered.}
\label{tab:overview_effects}
\begin{tabular}{lrrr}
  \hline
 & \thead{\textbf{BIC}} & \thead{\textbf{AIC}} & \thead{\textbf{BIC} \\ \textbf{relaxed}}\\
  \hline
Jean-Marie Pfaff (goalkeeper) & 0.000 & -0.001 & -0.125 \\
  Rudolf Kargus (goalkeeper) & 0.000 & 0.000 & 0.000\\
  $\vdots$ & $\vdots$ & $\vdots$ & $\vdots$ \\
  Manuel Neuer (goalkeeper) & 0.000 & 0.000 & 0.000\\ 
  Lothar Matth\"aus (player) & 0.000 & 0.000 & 0.000\\ 
    $\vdots$ & $\vdots$ & $\vdots$ & $\vdots$ \\
  Nuri Sahin (player) & 0.000 & 0.000 & 0.000\\ 
   \hline
\end{tabular}
\end{table}

\FloatBarrier
\section{Discussion}\label{chap:concl}
The modelling framework developed in this contribution, a LASSO-HMM and a relaxed-LASSO-HMM, respectively, allows for implicit variable selection in the state-dependent process of HMMs. The performance of the variable selection is first investigated in a simulation study, indicating that the relaxed-LASSO-HMM with the corresponding tuning parameter selected by the BIC is the best-performing fitting scheme considered. 

For the analysis of a hot shoe effect, we fit both LASSO-HMMs and relaxed-LASSO-HMMs to data on penalty kicks in the German Bundesliga. Factors potentially affecting the performance of penalty-takers, such as the current score of the match, are included in the predictors. In addition, dummy variables for the penalty-takers as well as for the goalkeepers are included. The results indicate evidence for a hot shoe effect. In addition, our results also shed some light on exceptionally performing players such as Jean-Marie Pfaff, a former goalkeeper of Bayern Munich, who has been selected by several of the considered fitting schemes.

A clear limitation of the real data application considered is the problem of self selection. Since the manager (or the team) can decide which player has to take the penalty, players who have been rather unsuccessful in the past may not take penalty kicks anymore. However, several teams have demonstrated in the past that they rely on and trust in a certain player for taking penalty kicks, regardless of the outcome of the kick. Whereas penalty kicks in football yield to a time series due to the way in which penalties take place, the corresponding time intervals between actions are irregular. Although our data cover all attempts in the German Bundesliga, there are sometimes several month between two attempts. Moreover, some players might be involved in penalty kicks in matches from other competitions such as the UEFA Champions League, the UEFA European Cup or in matches with their national teams. These could also affect, whether a player is currently in a hot or cold state. From this perspective, the time series of Bundesliga penalties could be considered as partly incomplete for some players.

From a methodological point, the number of states selected (i.e., $N=2$) may be too coarse for modelling the underlying form of a player. Considering a continuously varying underlying state variable instead may be more realistic, since gradual changes in a player's form could then be captured. This could be achieved by considering state-space models, where regularized estimation approaches are a first point for further research. The motivation in this contribution for $N = 2$ states, however, was to approximate the potential psychological states in a simple manner for ease of interpretation, e.g., in the sense of hot (``player is confident'') or cold (``player is nervous'') states. Moreover, our main focus was to show the usefulness of our method developed in a rather simple/intuitive setting with two states. In addition, future research could focus on regularization in HMMs where not only the intercept (as considered here), but also the parameters $\beta_j$ are allowed to depend on the current state. Regularization in this model formulation could be taken into account by applying so-called fused LASSO techniques (see, e.g., \citealp{Lasso:GerTut:2010}), where the parameters could either be shrunk to zero or to the same size for all states considered. For the former, a single tuning parameter for each state has to be introduced, which brings the need for efficient tuning strategies.

The modeling framework developed here can easily be tailored to other applications, where implicit variable selection in HMMs is desired. For the application considered in this contribution, i.e.\ an analysis of a potential hot hand/hot shoe effect, other sports such as basketball or hockey could be analyzed. Potential covariates --- whose corresponding effects are penalized ---  in these sports cover the shot type, shot origin, and game score, to name but a few.

\subsection*{Acknowledgements}
We want to thank the group of researchers B. Bornkamp, A. Fritsch, L. Geppert, P. Gn\"andinger, K. Ickstadt and O. Kuss
for providing the German Bundesliga penalty data set.

\newpage

\FloatBarrier
\bibliographystyle{apalike}
\bibliography{refs}

\end{spacing}
\end{document}